\documentclass[runningheads]{llncs}
\usepackage{graphicx}
\usepackage{hyperref}
\usepackage{amsmath}
\usepackage{amssymb}
\usepackage{makecell}
\usepackage{pgfplots}
\usepackage{bm}
\usepackage{hyphenat}
\usepackage{pgfplotstable}
\usepackage{float}
\usepackage{multirow}
\usepackage{caption}
\usepackage{subcaption}
\usepackage[export]{adjustbox}

\usetikzlibrary{matrix,calc,shapes,arrows}

\begin{document}
\title{Learning Sparse Masks for \\Diffusion-based Image Inpainting\thanks{This 
work has received funding from the European Research Council 
              (ERC) under the European Union's Horizon 2020 research and 
              innovation programme (grant agreement no. 741215, ERC Advanced 
              Grant INCOVID).}}
%
%
\author{Tobias Alt \and
        Pascal Peter \and
        Joachim Weickert}
\authorrunning{T. Alt et al.}
%
\institute{Mathematical Image Analysis Group,\\
         Faculty of Mathematics and Computer Science,\\
         Campus E1.7, Saarland University, 66041 Saarbr\"ucken, Germany.\\
         \email{\{alt, peter, weickert\}@mia.uni-saarland.de}}
\maketitle              
\begin{abstract}
Diffusion-based inpainting is a powerful tool for the reconstruction of images 
from sparse data. Its quality strongly depends on the choice of 
known data. Optimising their spatial location -- the inpainting mask -- is 
challenging. A commonly used tool for this task are stochastic optimisation 
strategies. However, they are slow as they compute multiple inpainting results. 
We provide a remedy in terms of a learned mask generation model. By emulating 
the complete inpainting pipeline with two networks for mask generation and 
neural surrogate inpainting, we obtain a model for highly efficient adaptive 
mask generation. Experiments indicate that our model can achieve competitive 
quality with an acceleration by as much as four orders of magnitude. Our 
findings serve as a basis for making diffusion-based inpainting more attractive 
for applications such as image compression, where fast encoding is 
highly desirable.
\end{abstract}
\section{Introduction}\label{sec:intro}
Inpainting is the task of restoring an image from limited amounts of data. 
Diffusion processes are particularly powerful for reconstructions from 
sparse data; see e.g.~\cite{WW06}. By solving a partial 
differential equation (PDE), they propagate information from a small known 
subset of pixels, the inpainting mask, to the missing image areas. Inpainting 
from sparse data is successful in applications such as image 
compression~\cite{GWWB08,PHNH16,SPME14}, adaptive sampling~\cite{DCPC19}, and 
denoising~\cite{APW17}.

Optimising the inpainting mask is essential for a good reconstruction. 
However, this is a challenging combinatorial problem. While there are 
theoretical results on optimal masks~\cite{BBBW08}, practical implementations 
are often qualitatively not that convincing albeit highly efficient. On the 
other hand, stochastic mask optimisation strategies~\cite{HMHW17,MHWT12} 
produce high quality masks, but are computationally expensive.

In the present paper, we combine efficiency and quality of mask optimisation 
for PDE-based inpainting with the help of deep learning. To this end, we design 
a hybrid architecture which, to the best of our knowledge, constitutes the 
first instance of learned sparse masks for PDE-based inpainting. 


\subsubsection{Our Contribution.}
We present a model for learning sparse inpainting masks for 
homogeneous diffusion inpainting. This type of inpainting shows good 
performance for optimised masks~\cite{MHWT12}, and does not depend on any free 
parameters. We employ two networks: one which generates 
a sparse inpainting mask, and one which acts as a surrogate solver for 
homogeneous diffusion inpainting. By using different loss functions for the two 
networks, we optimise both inpainting quality and fidelity to the inpainting 
equation.

The use of a surrogate solver is a crucial novelty in our work. It 
reproduces results of a diffusion-based inpainting process without having to 
perform backpropagation through iterations of a numerical solver. This 
replicates the full inpainting pipeline to efficiently 
train a mask optimisation model. 

We then evaluate the quality of the learned masks in a learning-free inpainting 
setting. Our model combines the speed of instantaneous mask generation 
approaches~\cite{BBBW08} with the quality of stochastic 
optimisation~\cite{MHWT12}. Thus, we reach a new level in sparse mask 
optimisation for diffusion-based inpainting. 


\subsubsection{Related Work.}
Diffusion-based inpainting plays a vital role in 
image and video compression~\cite{APMW21,GWWB08,SPME14}, 
denoising~\cite{APW17}, and many more. 
A good inpainting mask is crucial for successful image inpainting. 
Current approaches for the spatial optimisation of sparse
inpainting data in images can be classified in four categories.

\begin{enumerate}
  \item \emph{Analytic Approaches.} Belhachmi et 
  al.~\cite{BBBW08} have shown that in the continuous setting, optimal masks 
  for homogeneous diffusion inpainting can be obtained from the Laplacian 
  magnitude of the image. In practice this strategy is very fast, allowing 
  real-time inpainting mask generation by dithering the Laplacian magnitude. 
  However, the reconstruction quality is lacking, mainly due to limitations in 
  the quality of the dithering operators~\cite{HMHW17,MHWT12}.
  
  \item \emph{Nonsmooth Optimisation Strategies.} Several 
  works~\cite{BLPP16,CRP14,HMHW17,OCBP14} consider sophisticated nonsmooth 
  optimisation approaches that offer high quality, but do not allow to specify 
  the desired mask density in advance. Instead one influences it by varying a 
  regularisation parameter, which requires multiple program runs, resulting in 
  a slow runtime. Moreover, adapting the model to different inpainting 
  approaches is not trivial.

  \item \emph{Sparsification Methods.} They successively remove 
  pixel data from the image to create an adaptive inpainting mask. For example, 
  the \emph{probabilistic sparsification (PS)} of Mainberger et 
  al.~\cite{MHWT12} randomly removes a set of points and reintroduces a 
  fraction of those points with a high inpainting error. Sparsification 
  strategies are generic as they work with various inpainting operators such as 
  diffusion-based ones~\cite{HMHW17,MHWT12} or interpolation on 
  triangulations~\cite{DDI06,MMCB18}. Moreover, they allow to specify the 
  desired mask density in advance. However, they are also
  computationally expensive as they require many inpaintings to judge the 
  importance of individual data points to the reconstruction. Due to their 
  simplicity and their broad applicability, sparsification approaches are the 
  most widely used mask optimisation strategies. 
 
  \item \emph{Densification Approaches.} Densification 
  strategies~\cite{CW21,DAW21,KBPW18} start with empty or very sparse masks 
  and successively populate them. This makes them reasonably efficient, while 
  also yielding good quality. They are 
  fairly easy to implement and work well for PDE-based~\cite{CW21,DAW21} and 
  exemplar-based~\cite{KBPW18} inpainting operators. Still, they require 
  multiple inpainting steps in the range of 10 to 100 to obtain a sufficiently 
  good inpainting mask. 
\end{enumerate}

In order to escape from suboptimal local minima, the Categories 3 and 4 have 
been improved by \emph{nonlocal pixel exchange} (NLPE)~\cite{MHWT12}, at the 
expense of additional inpaintings and runtime. Moreover, it is well-known that 
optimising the grey or colour values of the mask pixels -- so-called tonal 
optimisation -- can boost the quality even further~\cite{HMHW17,MHWT12}. Also 
the approaches of Category 2 may involve tonal optimisation implicitly or 
explicitly.

Qualitatively, carefully tuned approaches of Categories 2--4 play in a
similar league, and are clearly ahead of Category 1. However, their runtime is
also substantially larger than Category 1, mainly due to the many 
inpaintings that they require. Last but not least, all aforementioned
approaches are fully model-based, in contrast to most recent approaches 
in image analysis that benefit from deep learning ideas.

The goal of the present paper is to show that the incorporation of
deep learning can give us the best of two worlds: a real-time capability
similar to Category~1, and a quality similar to Categories 2--4. In order to 
focus on the main ideas and to keep things simple, we restrict ourselves to 
homogeneous diffusion inpainting and compare only to probabilistic 
sparsification without and with NLPE. Also tonal optimisation is 
not considered in our paper, but is equally possible for our novel
approach. More refined approaches and more comprehensive evaluations will be 
presented in our future work.

Learning-based inpainting has also been successful in recent years. Following 
the popular work of Xie et al.~\cite{XXC12}, several architectures and 
strategies for inpainting have been  
proposed; see e.g.~\cite{IIST17,LJXY19,PKDD16,YLLS17,YLYS18}. 
However, inpainting from sparse data is rarely considered. Va\v{s}ata et
al.~\cite{VHF21} present sparse inpainting based on Wasserstein generative 
adversarial networks. Similarly, Ulyanov et al.~\cite{UVL18} consider 
inpainting from sparse data without mask generation. Dai et al.~\cite{DCPC19} 
present a trainable mask generation model from an adaptive sampling 
viewpoint. Our approach is the first to combine deep learning for mask 
optimisation for PDE-based inpainting in a transparent and efficient 
way.


\subsubsection{Organisation of the Paper.}
In Section \ref{sec:review}, we briefly review diffusion-based inpainting. 
Afterwards in Section \ref{sec:ours}, we introduce our model for 
learning inpainting masks. We evaluate the quality of the learned masks in 
Section \ref{sec:experiments} before presenting our conclusions in Section 
\ref{sec:conclusion}.
 

\section{Review: Diffusion-based Inpainting}\label{sec:review}
The goal of inpainting is to restore missing information in a continuous 
greyscale image 
$f:\Omega\rightarrow\mathbb R$ on some rectangular domain $\Omega$, where 
image data is only available on an inpainting mask $K \subset \Omega$. In this 
work we focus on homogeneous diffusion inpainting, which computes the 
reconstruction $u$ as the solution of the PDE
\begin{equation}\label{eq:inp}
  \left(1 - c\right) \Delta u - c \left(u - f\right) = 0
\end{equation}
with reflecting boundary conditions. Here, a confidence measure 
$c:\Omega\rightarrow\mathbb R$ denotes whether a value is known or not. Most 
diffusion-based inpainting models consider binary values for $c$: A value of 
$c(\bm x)=1$ indicates known data and thus $u=f$ on $K$, while $c(\bm x)=0$ 
denotes missing data, leading to homogeneous diffusion~\cite{Ii62} inpainting 
$\Delta u=0$ on $\Omega\backslash K$, where $\Delta = \partial_{xx} + 
\partial_{yy}$ denotes the Laplacian. However, it is also possible to use 
non-binary confidence measures~\cite{HW15}, which we will exploit to our 
advantage.

We consider digital greyscale images $\bm u, \bm f \in 
\mathbb{R}^{n_xn_y}$ with dimensions $n_x \times n_y$ and discretise the 
inpainting equation \eqref{eq:inp} by means of finite differences. Then a 
numerical solver for the resulting linear system of equations is used to obtain 
a reconstruction $\bm u$. For a good inpainting quality, optimising the binary 
mask $\bm c \in \{0,1\}^{n_xn_y}$ is crucial. This problem is constrained by a 
desired mask density $d$ which measures the percentage of mask pixels 
w.r.t.~the number of image pixels. 

One strategy for mask optimisation has been proposed by Belhachmi et 
al.~\cite{BBBW08}. They show that an optimal mask in the continuous setting can 
be obtained from the rescaled Laplacian magnitude of the image. However, 
transferring these results to the discrete setting often suffers from 
suboptimal dithering strategies. While being highly efficient, reconstruction 
quality is not fully satisfying.

Better quality can be achieved with the popular stochastic 
strategies of Mainberger et al.~\cite{MHWT12}. First, one employs 
\emph{probabilistic sparsification (PS)}: Starting with a full mask, one 
removes a fraction~$p$ of candidate pixels and computes the inpainting. Then 
one reintroduces a fraction~$q$ of the candidates with the largest local 
inpainting error. One repeats this step until reaching a desired mask 
density~$d$.

Since sparsification is a greedy local approach, it can get trapped in bad 
local minima. As a remedy, Mainberger et al.~\cite{MHWT12} also propose a 
\emph{nonlocal pixel exchange (NLPE)}. Pixel candidates in a sparsified mask 
are exchanged for an equally large set of non-mask pixels. If the new 
inpainting result improves, the exchange is kept, otherwise it is discarded. In 
theory, NLPE can only improve the mask, but in practice convergence is slow. 

The use of PS and NLPE requires to solve the inpainting problem numerous 
times, leading to slow mask optimisation. To avoid this computational 
bottleneck, we want to reach the quality of stochastic mask optimisation with a 
more efficient model based on deep learning.


\begin{figure*}[t]
  \centering
  \resizebox{\linewidth}{!}{\input{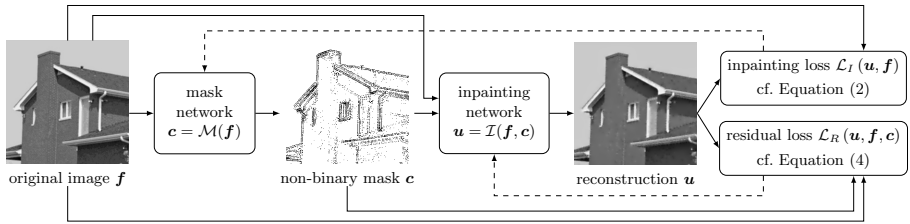}}
  \caption{Overview over our model structure. Solid lines denote forward 
  passes, dashed lines denote backpropagation. \label{fig:model_structure}}
\end{figure*}

\section{Sparse Masks with Surrogate Inpainting}\label{sec:ours}
Our model consists of two equally shaped U-nets~\cite{RFB15} with different 
loss functions. By optimising both inpainting quality and fidelity to the 
inpainting equation, we obtain masks with good reconstruction quality for the 
inpainting problem.


\subsection{The Mask Network}
The \emph{mask network} takes an original image $\bm f$ and transforms it into 
a mask $\bm c$. We denote the forward pass through the mask network by 
$\mathcal{M}(\cdot)$, i.e. the mask is computed as $\bm c = \mathcal{M}(\bm 
f)$. 

The mask entries lie in the interval $\left[0,1\right]$. Permitting non-binary 
values allows for a differentiable network model. To obtain mask points in the 
desired range, we apply a sigmoid function to the output of the network. 
Moreover, the mask network is trained for a specific mask density $d$. To this 
end, we rescale the outputs of the network if they exceed the desired density. 
We do not require a lower bound, since the loss function incites a sufficiently 
dense mask. 

The mask network places the known data such that the inpainting error between 
the reconstruction $\bm u$ and the original image $\bm f$ is minimised. This 
yields the \emph{inpainting loss}
\begin{equation}\label{eq:inploss}
  \mathcal{L}_I \!\left(\bm u, \bm f\right) = \frac{1}{n_xn_y} \|\bm u -\bm 
  f\|_2^2
\end{equation}
as its objective function where $\|\cdot\|_2$ is the Euclidean norm. Its 
implicit dependency on the inpainting mask links the learned masks to the 
reconstructions.

We found that the mask network tends to get stuck in local minima with flat 
masks which are constant at every position, yielding a random sampling.  
To avoid this, we add a regularisation term $\mathcal R(\bm c)$ 
to the inpainting loss $\mathcal L_I(\bm u, \bm f)$. It penalises 
the inverse variance of the mask via $\mathcal{R}(\bm c) = \left(\sigma^2_{\bm 
c} + \epsilon\right)^{-1}$ where a small constant $\epsilon$ avoids 
division by zero. The variance of a mask describes how strongly the confidence 
measures of the individual pixels differ from the mean probability. Thus, the 
regulariser serves two purposes: First, it lifts the bad local minima for flat 
masks by adding a strong penalty to the energy. Second, it promotes 
probabilities closer to $0$ and $1$, as this maximises the variance. The 
impact of the regularisation term is steered by a positive regularisation 
parameter~$\alpha$. 


\subsection{The Inpainting Network}
The second network is called the \emph{inpainting network}. Its task is to 
create a reconstruction $\bm u$ which follows a classical inpainting process. 
In~\cite{ASAPW21}, it has been shown that U-nets realise an efficient multigrid 
strategy at their core. Thus, we use a U-net as a surrogate solver which 
reproduces the results of the PDE-based inpainting. The inpainting network 
takes the original image $\bm f$ and the mask $\bm c$ and creates a 
reconstruction $\bm u = \mathcal{I}\left(\bm f, \bm c\right)$. This result 
should solve the discrete version of the inpainting equation \eqref{eq:inp} 
which reads
\begin{equation}
  \left(\bm I - \bm C\right) \bm A \bm u - \bm C \left(\bm u -\bm 
    f\right) = 0.
\end{equation}
Here, $\bm A$ is a discrete implementation of the Laplacian $\Delta$ 
with reflecting boundary conditions, and $\bm C = \text{diag}(\bm c)$ is a 
matrix representation of the mask. To ensure that the reconstruction $\bm u$ 
approximates a solution to this equation, we minimise its residual, yielding 
the \emph{residual loss}
\begin{equation}\label{eq:resloss}
  \mathcal{L}_R \!\left(\bm u, \bm f, \bm c\right) = \frac{1}{n_xn_y} 
  \| \left(\bm I - \bm C\right) \bm A \bm u - \bm C \left(\bm u -\bm 
  f\right)\|_2^2 \,.
\end{equation}
As the residual loss measures fidelity to the PDE-based process, an optimal 
network approximates the PDE solution in an efficient way that allows fast 
backpropagation. This strategy has been proposed in~\cite{ASAPW21} and is 
closely related to the idea of deep energies~\cite{GFE21}.

Figure \ref{fig:model_structure} presents an overview of the full model 
structure. Note that the inpainting network receives both the mask and the 
original image as an input. Thus, this network is not designed for standalone 
inpainting. However, this allows the network to easily minimise the residual 
loss by transforming the original into an accurate inpainting result, given the 
mask as side information.


\subsection{Practical Application}
After training the full pipeline in a joint fashion, the mask network can be 
used to generate masks for homogeneous diffusion inpainting. To this end, 
we apply the mask network to an original image and obtain a non-binary 
mask. This mask is then binarised: The probability of a pixel 
belonging to a mask is given by its non-binary value. At each position, we 
perform a weighted coin flip with that probability. Afterwards, the 
binarised masks are fed into a numerical solver of choice for homogeneous 
diffusion inpainting. 

While binarising the mask is not necessary in this pure inpainting 
framework, it is important for compression applications since storing binary 
masks with arbitrary point distributions is already highly 
non-trivial~\cite{MPW21}.


\begin{figure}[t]
  \centering 
  \setlength{\tabcolsep}{2pt}
  \begin{tabular}{ccccc}
    \includegraphics[width=0.18\linewidth]{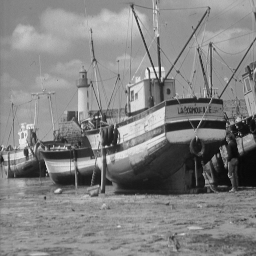}
   &\includegraphics[width=0.18\linewidth]{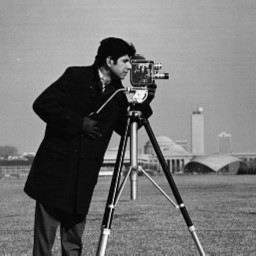}
   &\includegraphics[width=0.18\linewidth]{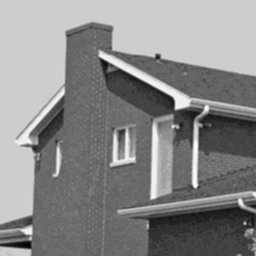}
   &\includegraphics[width=0.18\linewidth]{figures/trui}
   &\includegraphics[width=0.18\linewidth]{figures/peppers}
   \\
   \emph{boat}
   &\emph{cameraman}
   &\emph{house}
   &\emph{trui}
   &\emph{peppers}
  \end{tabular}
  \caption{Test images of resolution $256\times256$.\label{fig:images}}
\end{figure}

\section{Experiments}\label{sec:experiments}


\subsection{Experimental Setup}
We train both U-nets jointly with their respective loss function on the BSDS500 
dataset~\cite{AMFM11} which contains a broad selection of natural images. 
As a training set, we use $200$ cropped grey value 
images of size $256\times256$ with values in the range $[0,255]$. We do 
not use a validation set as the training process is fully fixed.

Both U-nets employ $5$ scales, with $3$ layers per scale. On the finest scale, 
they use $10$ channels, and this number is doubled on each scale. Thus, each 
U-net possesses around $9\times10^5$ parameters. We use the Adam 
optimiser~\cite{KB15} with standard settings, a learning rate of $5 \cdot 
10^{-4}$, and $4000$ epochs. As a regularisation parameter we choose $\alpha 
=0.01$. We found this combination of hyperparameters to work well in 
practice. We train multiple instances of the model for densities 
between~$10\%$ and~$1\%$ with several random initialisations. 

After training, we binarise the masks and use them with a conjugate gradient
solver for homogeneous diffusion inpainting to obtain a reconstruction. Since 
we aim at the highest quality, we take the best result out of $30$ samplings.

\begin{figure}[hp]
  \centering 
  \setlength{\tabcolsep}{1pt}
  \begin{subfigure}{\textwidth}
  \begin{tabular}{cccc}
    original
  & Belhachmi et al.
  & PS + NLPE
  & our model
  \\
    \includegraphics[width=0.24\linewidth]{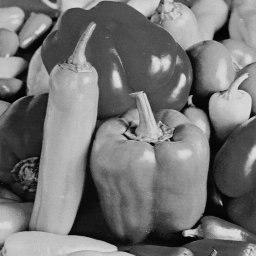}
  & \includegraphics[width=0.24\linewidth]
         {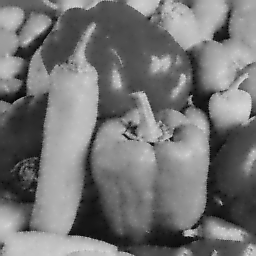}
  & \includegraphics[width=0.24\linewidth]
      {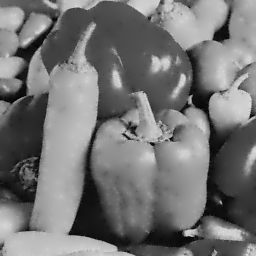}
  & \includegraphics[width=0.24\linewidth]
      {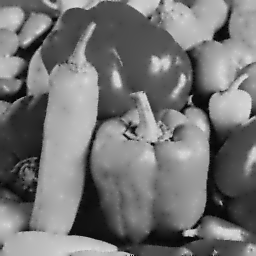}
  \\
  & \includegraphics[width=0.24\linewidth,frame]
      {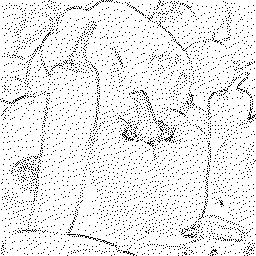}
  & \includegraphics[width=0.24\linewidth,frame]
      {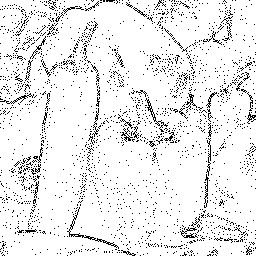}
  & \includegraphics[width=0.24\linewidth,frame]
      {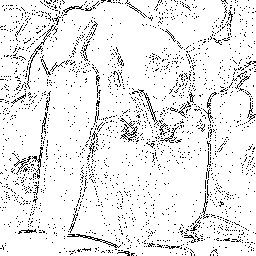}
  \end{tabular}
  \caption{\emph{peppers} with $8\%$ density}
  \vspace{5mm}
  \end{subfigure}
  \begin{subfigure}{\textwidth}
  \begin{tabular}{cccc}
  original
  & Belhachmi et al.
  & PS + NLPE
  & our model
  \\
    \includegraphics[width=0.24\linewidth]{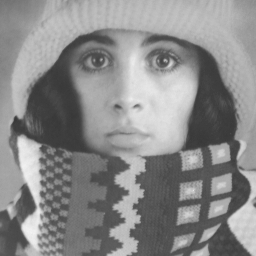}
  & \includegraphics[width=0.24\linewidth]
         {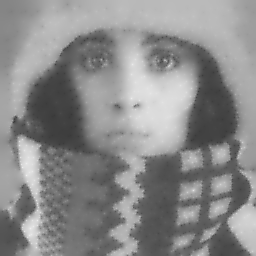}
  & \includegraphics[width=0.24\linewidth]
       {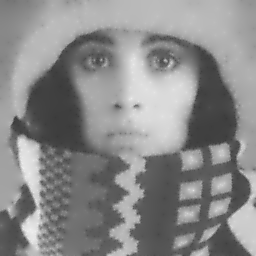}
  & \includegraphics[width=0.24\linewidth]
       {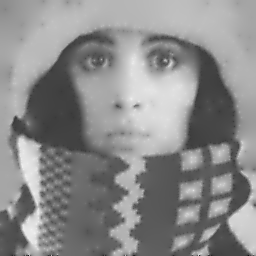}
  \\
  & \includegraphics[width=0.24\linewidth,frame]
       {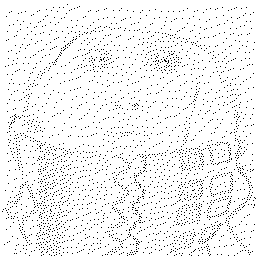}
  & \includegraphics[width=0.24\linewidth,frame]
     {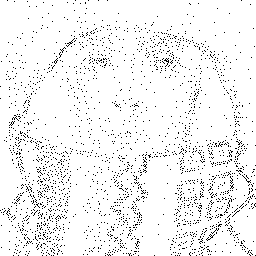}
  & \includegraphics[width=0.24\linewidth,frame]
     {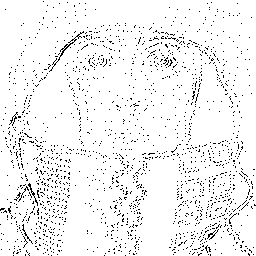}
  \end{tabular}
  \caption{\emph{trui} with $5\%$ density}
  \vspace{3mm}
  \end{subfigure}
    \caption{Visual comparison of inpainting results on two exemplary
    images for different mask densities. Mask points are shown in black, and 
    mask images are framed for better visibility. Top rows depict the 
    inpainting results, and bottom rows display the masks, respectively. The 
    learned masks yield inpainting results which are visually comparable to PS 
    with additional NLPE.    
    \label{fig:quality}}
\end{figure} 

Analogously, we generate masks with PS as well as with PS with additional NLPE. 
In the following, we denote the latter combination by PS+NLPE. In our 
sparsification we use candidate fractions $p=0.1$ and $q=0.05$ as 
suggested by Mainberger et al.~\cite{MHWT12}, and we take the 
best result out of $5$ runs. For NLPE, we use $30$ candidates of which $10$ are 
exchanged. We run NLPE for $10$ cycles: In a single cycle, each mask point is 
exchanged once on average. Moreover, we compare against the strategy of 
Belhachmi et al.~\cite{BBBW08}. This approach is realised by taking the 
Laplacian magnitude of the image, rescaling it to obtain a desired density, and 
dithering the result with a binary Floyd--Steinberg algorithm~\cite{FS76}.

We compare our results on five popular test images (see 
Figure~\ref{fig:images}), since performing PS and 
NLPE on a large database is infeasible. We measure the quality in terms 
of peak signal-to-noise ratio (PSNR). Higher values indicate better 
quality. 

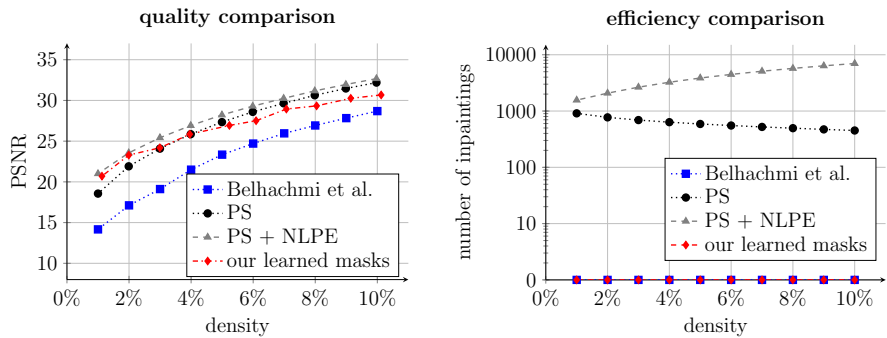
\begin{figure}[t]
  \centering 
  \resizebox{0.48\linewidth}{!}{\def\FILE{figures/quality_results}

\def\PS{0}
\def\NLPE{1}
\def\CNN{2}
\def\BEL{3}

\pgfplotstablesort[sort key=density, sort cmp=float >]{\tablesorted}{\FILE}

\begin{tikzpicture}[scale = 1.0]
\begin{axis}
[
width=\axisdefaultwidth,
height=0.75*\axisdefaultwidth,
samples=500,
axis lines=left,
ymin = 8, ymax = 37,
xmin = 0.0, xmax = 0.11, 
ytick = {0,10,15,...,35},
xtick = {0, 0.02, 0.04, 0.06, 0.08, 0.1},
xticklabels={0\%, 2\%, 4\%, 6\%, 8\%, 10\%},
ylabel={PSNR},
xlabel={density},
xlabel near ticks,
ylabel near ticks,
grid = major,
legend entries={Belhachmi et al.,
                PS,
                PS + NLPE,
                our learned masks},
legend cell align=left,
legend style={at={(0.35, 0.23)}, anchor = west},
font=\large,
title={\textbf{quality comparison}}
]

\addplot+[thick, dotted, mark options={solid}, blue, mark = square*,
restrict expr to domain={\thisrow{approachid}}{\BEL:\BEL},
unbounded coords=discard]
table[x = density, y expr= 10*log10(255*255/\thisrow{mse})]
{\tablesorted};

\addplot+[thick, dotted, mark options={solid}, black, mark = *,
restrict expr to domain={\thisrow{approachid}}{\PS:\PS},
unbounded coords=discard]
table[x = density, y expr= 10*log10(255*255/\thisrow{mse})]
{\tablesorted};

\addplot+[thick, dashed, mark options={solid}, gray, mark = triangle*,
restrict expr to domain={\thisrow{approachid}}{\NLPE:\NLPE},
unbounded coords=discard]
table[x = density, y expr= 10*log10(255*255/\thisrow{mse})]
{\tablesorted};

\addplot+[thick, dashdotted, mark options={solid}, red, mark = diamond*,
restrict expr to domain={\thisrow{approachid}}{\CNN:\CNN},
unbounded coords=discard]
table[x = density, y expr= 10*log10(255*255/\thisrow{mse})]
{\tablesorted};

\end{axis}
\end{tikzpicture}} 
  \hfill
  \resizebox{0.51\linewidth}{!}{\def\FILE{figures/speed_results}

\def\PS{0}
\def\NLPE{1}
\def\CNN{2}
\def\BEL{3}

\begin{tikzpicture}[scale = 1.0]
\begin{axis}
[
width=\axisdefaultwidth,
height=0.75*\axisdefaultwidth,
samples=500,
axis lines=left,
ymin = 0, ymax = 15000, ymode=log,
xmin = 0.0, xmax = 0.11, 
ytick = {1,10,100,1000,10000},
yticklabels = {0,10,100,1000,10000},
xtick = {0, 0.02, 0.04, 0.06, 0.08, 0.1},
xticklabels={0\%, 2\%, 4\%, 6\%, 8\%, 10\%},
ylabel={number of inpaintings},
xlabel={density},
xlabel near ticks,
ylabel near ticks,
grid = major,
legend entries={Belhachmi et al.,
                PS,
                PS + NLPE,
                our learned masks},
legend cell align=left,
legend style={at={(0.35, 0.3)}, anchor = west},
font=\large,
title={\textbf{efficiency comparison}}
]

\addplot+[thick, dotted, mark options={solid}, blue, mark = square*,
restrict expr to domain={\thisrow{approachid}}{\BEL:\BEL},
unbounded coords=discard]
table[x = density, y = inpaintings]
{\FILE};

\addplot+[thick, dotted, mark options={solid}, black, mark = *,
restrict expr to domain={\thisrow{approachid}}{\PS:\PS},
unbounded coords=discard]
table[x = density, y = inpaintings]
{\FILE};

\addplot+[thick, dashed, mark options={solid}, gray, mark = triangle*,
restrict expr to domain={\thisrow{approachid}}{\NLPE:\NLPE},
unbounded coords=discard]
table[x = density, y = inpaintings]
{\FILE};

\addplot+[thick, dashed, mark options={solid}, red, mark = diamond*,
restrict expr to domain={\thisrow{approachid}}{\CNN:\CNN},
unbounded coords=discard]
table[x = density, y = inpaintings]
{\FILE};

\end{axis}
\end{tikzpicture}} 
  \caption{Comparison of models in terms of quality and efficiency. 
  \textbf{(a) Left:} Average inpainting quality in PSNR for each density. 
  \textbf{(b) Right:} Efficiency in terms of the number of inpaintings for each 
  density. The learned masks consistently outperform those of Belhachmi et al. 
  and  can compete with masks generated by PS. For very sparse masks, our model 
  can compete with PS+NLPE. Both our method and that of Belhachmi et al. 
  generate masks without computing an inpainting. The stochastic optimisation 
  strategies compute up to thousands of inpaintings. 
  \label{fig:comparisons}}
\end{figure}

\subsection{Reconstruction Quality}
Figure \ref{fig:quality} shows a visual comparison of optimised masks and the 
corresponding inpainting results. For both test cases, we observe that our 
learned masks are structurally similar to those obtained by PS with NLPE. This 
helps to create sharper contours, whereas the inpainting results of Belhachmi 
et al. suffer from fuzzy edges. The visual quality of the inpainting results 
for our model and PS+NLPE is indeed competitive. 

Figure \ref{fig:comparisons}(a) presents a comparison of the reconstruction 
quality averaged over the test images. Our learned masks consequently 
outperform the strategy of Belhachmi et al.. Moreover, our model is on par with 
PS for densities smaller than $5\%$. For extremely small densities up to $2\%$, 
it even outperforms PS and is on par with PS+NLPE.

For larger mask densities, the margin between the methods becomes smaller, 
and our model cannot outperform its stochastic counterparts. Still, all 
models produce a good reconstruction quality. However, for applications 
such as inpainting-based image compression, very sparse masks are more 
important and more challenging~\cite{MHWT12,SPME14}. Therefore, our mask 
generation model performs well for the practically relevant mask densities.


\subsection{Computational Efficiency}
The decisive advantage of the learned mask generation is its speed. 
As inpainting operations are the dominant factor for computation time, we use 
the number of inpaintings as a measure for efficiency. In comparison, the 
forward pass of the mask network is negligible.

Figure \ref{fig:comparisons}(b) visualises the average number of inpaintings 
required to obtain masks of a specific density for the test set. To generate a 
mask, both our model and that of Belhachmi et al. do not require any inpainting 
operations. Thus, the efficiency of these mask generation strategies does not 
depend on the density.

For PS, lower densities require more inpainting 
operations. Adding NLPE requires even more inpaintings 
depending on the number of cycles and the mask density. Both 
strategies trade computational efficiency for inpainting quality. 

For example, a single sparsification run for a $3\%$ mask on the 
\emph{cameraman} image with realistic parameter settings requires $700$ 
steps. On an \emph{Intel Core i7-7700K CPU @ 4.20GHz}, this amounts to 
$58$ seconds of runtime. The subsequent NLPE optimisation requires another 
$2000$ steps, resulting in more than $3$ minutes of additional runtime. 
In contrast, the strategy of Belhachmi et al. does not require any 
inpainting, and a mask can be generated in only $24$ milliseconds. 

Our model requires only $85$ milliseconds for passing a single image through 
the mask network on the CPU. Thus, it plays in the same league as the strategy 
of Belhachmi et al., while being on par with the stochastic optimisation in 
terms of quality. This allows instantaneous high quality mask generation. As a 
consequence, our learned model can serve as a highly efficient replacement of 
stochastic mask optimisation.

\section{Conclusions}\label{sec:conclusion}
We have proposed the first approach of sparse mask learning for
diffusion-based inpainting. It fuses ideas from deep learning   
with classical homogeneous diffusion inpainting. The key of this
strategy is a combination of an inpainting loss for the mask
generator and a residual loss for the surrogate inpainting network.
Its results are competitive with stochastic mask optimisation, while
being up to four orders of magnitude faster. This constitutes a new
milestone in mask optimisation for diffusion-based inpainting.

We are currently extending this idea to more sophisticated inpainting
operators, as well as to further optimisations of the network
architecture. We hope that this will pave the way to overcome the 
current time-consuming data optimisation strategies and will 
become an essential component for real-time diffusion-based codecs  
in hitherto unmatched quality.

%
%
%
 \bibliographystyle{splncs04}
 \bibliography{myrefs}
\end{document}